\documentclass[
prl, 
twocolumn,
superscriptaddress, 
nofootinbib, 
amsmath, 
amssymb,
aps, 
floatfix
]{revtex4-2}
\usepackage{graphicx,xcolor,setspace}
\usepackage[
    colorlinks=true,
    linkcolor=blue,
    urlcolor=blue,
    citecolor=blue
]{hyperref}

\usepackage{bm}
\usepackage{mathtools}

\usepackage[capitalise]{cleveref}
\usepackage{siunitx}
\DeclareSIUnit{\year}{yr}

\begin{document}

\preprint{APS/123-QED}

\title{Impact of Dark Compton Scattering on Direct Dark Matter Absorption Searches}
\author{Yonit Hochberg}\email{yonit.hochberg@mail.huji.ac.il}
\affiliation{Racah Institute of Physics, Hebrew University of Jerusalem, Jerusalem 91904, Israel}

\author{Belina von Krosigk}%
\email{belina.krosigk@kit.edu}
\affiliation{Institute for Astroparticle Physics (IAP), Karlsruhe Institute of Technology (KIT), 76344 Eggenstein-Leopoldshafen, Germany}
\affiliation{
Institut f\"ur Experimentalphysik, Universit\"at Hamburg, 22761 Hamburg, Germany}

\author{Eric Kuflik}\email{eric.kuflik@mail.huji.ac.il}
\affiliation{Racah Institute of Physics, Hebrew University of Jerusalem, Jerusalem 91904, Israel}

\author{To Chin Yu}\email{ytc@stanford.edu}
\affiliation{Department of Physics, Stanford University, Stanford, California 94305, USA}
\affiliation{SLAC National Accelerator Laboratory, 2575 Sand Hill Road, Menlo Park, California 94025, USA}
\date\today

\begin{abstract}
Direct detection experiments are gaining in mass reach. Here we show that the inclusion of dark Compton scattering, which has typically been neglected in absorption searches, has a substantial impact on the reach and discovery potential of direct detection experiments at high bosonic cold dark matter masses. We demonstrate this for relic dark photons and axion-like particles: we improve expected reach across materials, and further use results from SuperCDMS, EDELWEISS and GERDA to place enhanced limits on dark matter parameter space. We outline the implications for detector design and analysis.
\end{abstract}

\maketitle

%%%%%%%%%%%%%
\section{Introduction}
The identity of dark matter (DM) is one of the biggest mysteries of modern day physics, and is a driving force for both theoretical and experimental activities in particle physics. A diverse suite of direct detection experiments are currently hunting for DM~\cite{XENON:2018voc,LUX:2017ree,SuperCDMS:2018mne,PandaX-II:2020oim,CRESST:2019jnq,EDELWEISS:2020fxc,DAMIC:2020cut,SENSEI:2020dpa,XMASS:2018bid,Baudis:2012bc,PICO:2019vsc,NEWS-G:2017pxg,DEAP:2019yzn,DarkSide:2018kuk,DAMA:2010gpn,COSINE-100:2019lgn,CDEX:2019isc,Amare:2019jul}, utilizing both scattering and absorption processes to set limits on dark matter parameter space. Many new ideas to probe dark matter in the laboratory have been proposed in recent years~\cite{Essig:2011nj,Graham:2012su,Essig:2015cda,Lee:2015qva,Hochberg:2015pha,Hochberg:2015fth,Alexander:2016aln,Derenzo:2016fse,Hochberg:2016ntt,Kavanagh:2016pyr,Emken:2017erx,Emken:2017qmp,Battaglieri:2017aum,Essig:2017kqs,Cavoto:2017otc,Hochberg:2017wce,Essig:2018tss,Emken:2018run,Ema:2018bih,Geilhufe:2018gry,Baxter:2019pnz,Essig:2019xkx,Emken:2019tni,Hochberg:2019cyy,Trickle:2019nya,Griffin:2019mvc,Coskuner:2019odd,Geilhufe:2019ndy,Catena:2019gfa,Blanco:2019lrf,Kurinsky:2019pgb,Kurinsky:2020dpb,Griffin:2020lgd,Radick:2020qip,Gelmini:2020xir,Trickle:2020oki,Du:2020ldo,Blanco:2021hlm}, adding to the wealth and strength of the DM direct detection program. As experiments gain in mass reach, the inclusion of additional interaction channels relevant to these masses that have thus far been neglected becomes an important task.

Here we focus on such interactions where a DM particle enters and interacts with the detector, and does not scatter out. Instead, it undergoes a Compton-like process, where the incoming bosonic DM particle interacts with the electrons in the detector and a photon is emitted in the final state. We identify this process as important when compared to high energy absorption of relic dark bosons, as its inclusion in experimental data analysis can significantly alter extracted limits, projected reach and discovery potential. 
%Absorption searches can probe relic dark boson masses as high as twice the electron mass and thus $\sim 1$\,MeV. Above this mass tree-level decays into electron-positron pairs become possible, resulting in a too short lifetime for dark bosons to be an adequate relic DM candidate.
%a process too fast for the dark bosons to survive until today and to be an adequate relic DM candidate.

This {\it Letter} is organized as follows. We first compute the rate for Compton-like processes that relic dark bosons can undergo. We then discuss detector considerations and present our results on dark matter parameter space, including the impact on zero-background projections across materials and on existing limits. We end with a discussion and outlook.

\begin{figure}[th!] 
    \centering
\includegraphics[trim={0cm .5cm 4.8cm 0},clip]{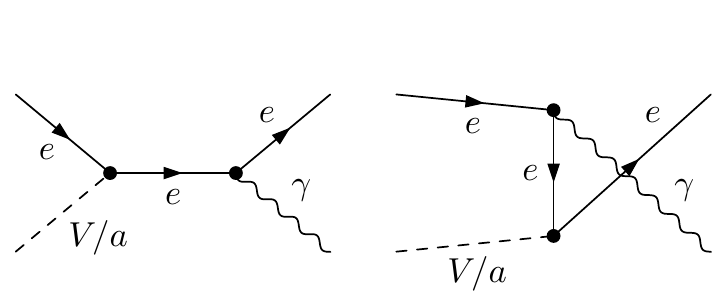}
\caption{
   Tree-level Feynman diagrams illustrating the Compton-like process where a dark matter boson is converted to a photon via electron scattering. 
    }
    \label{fig:feynman_diagram}
\end{figure}

%%%%%%%%%%%%%
\section{Dark Compton Scattering Rates}%
We begin by calculating the rates of dark Compton scattering for dark bosons, illustrated in Fig.~\ref{fig:feynman_diagram}. We compute the dark Compton scattering for dark photons and axion-like particles (ALPs) using the free-electron approximation (FEA). Details of the rate calculations can be found in the Supplemental Material \cite{SM}. The use of the FEA is justified here since the rate of dark Compton scattering will only be significant compared to the absorption rate of the dark bosons when the mass of DM is large ($m_{\rm DM}\gtrsim 100$\,keV). In this regime, the energy received by the electron in the event is larger than most atomic binding energies and can thus be treated as approximately free \cite{qiao2020compton}. We defer calculation of corrections due to core electron levels to future work.

\begin{figure*}[t!]
    \centering
    \includegraphics[width=0.9\textwidth]{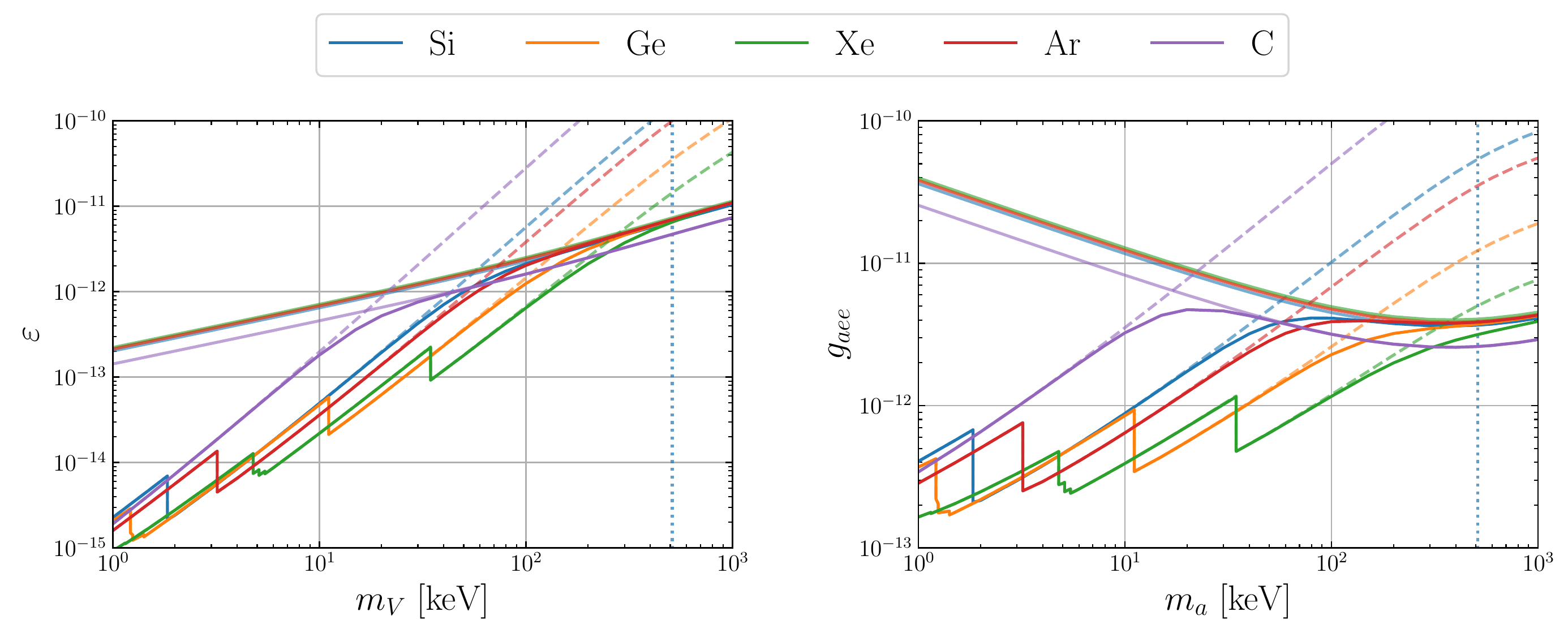}
    \caption{
    Projected 90\% C.L. reach for relic dark photons ({\it left}) and axion-like particles ({\it right}) with a 1 kg-year exposure and no background. The reach for absorption (dashed lines) and Compton-like processes (thin solid lines) is indicated along with the combined reach (thick solid lines), showing substantially improved capabilities at high DM masses when dark Compton scattering is included. The vertical dotted line indicates the mass of the electron.
    }
    \label{fig:zero_bkg_limits}
\end{figure*}

{\bf Dark Photons.} Consider a dark photon, $V$, of mass $m_V$ scattering with electrons and emitting a photon, ${e^- + V \to e^- + \gamma}$ 
(see Fig. \ref{fig:feynman_diagram}). The energy of the emitted photon is 
\begin{align}
    \omega' &= \frac{m_V^2+2m_e\omega}{2(m_e+\omega-\sqrt{\omega^2-m_V^2}\cos\theta)} \,,
\end{align}
where $\omega$ is the energy of the incoming dark photon, $m_e$ is the mass of the electron and $\theta$ is the angle of the emitted photon.

The dark Compton rate for a slow-moving DM flux with $\omega\approx m_V$ is given by
\begin{align}
    \label{eq:Rate_DP_long}
    R &= \frac{n_e}{\rho_T}\frac{\rho_{\rm DM}}{m_{V}} \frac{e^4\varepsilon^2}{24\pi m_e^2}\frac{ (m_V+2m_e) (m_V^2+2m_em_V+2m_e^2) }{(m_V + m_e)^3}, 
\end{align}
where $\varepsilon$ is  the  kinetic  mixing  strength  of  the  dark  photon, $\rho_T$ is the target density, and $n_e$ is the number density of electrons.  
Importantly, for $\omega=m_V$, the electron recoil energy is fixed to be $T=\omega-\omega'=\frac{m_V^2}{2(m_e+m_V)}$. The differential rate $dR/dT$ is then a delta function---just as is the case for an absorption process---however the value of the recoil energy is smaller than for the absorption case.

{\bf Axion-like Particles. }
Next consider an ALP, $a$, of mass $m_a$, that couples to the electron with strength $g_{aee}$ and undergoes a dark Compton scattering. For a slow-moving DM flux, the rate is given by 
\begin{align}
\label{eq:Rate_ALP_long}
R &= \frac{n_e \rho_{\rm DM}}{\rho_T}\frac{e^2 g^2_{aee} m_a (m_a+2m_e)^2}{ 16\pi m_e^2(m_a+m_e)^4}\,.
\end{align}
The electron recoil spectrum is again a delta function centered at the same value as for a dark photon, at ${T=\frac{m_a^2}{2(m_e+m_a)}}$.

{\bf Absorption.} For comparison, the rates for absorption of dark photons and ALPs are given by \cite{Hochberg:2016sqx,Bloch_2017} 
\begin{equation}
    \label{equ:AbsRate_DP}
    R_{{\rm abs},V} = \frac{\rho_\mathrm{DM}}{\; m_{V}} \varepsilon_{\mathrm{eff}}^2 \sigma_{\rm PE}(m_{V})\,
\end{equation}
 with $\varepsilon_{\mathrm{eff}}\approx \varepsilon$ for $m_V \gtrsim 20$\,eV \cite{Hochberg:2016sqx} 
and 
\begin{equation}
\label{equ:AbsRate_ALPs}
    R_{{\rm abs},a} = \rho_\mathrm{DM} \frac{3 g_{aee}^2 \, m_a}{4 \,e^2\, m_e^2}\sigma_{\rm PE}(m_a)\,  \,,
\end{equation}
 with $\sigma_{\rm PE}$ being the photoelectric absorption cross section~\cite{berger1987xcom} in the target material at the relevant energy.

%%%%%%%%%%%%%
\section{Detector considerations}

\begin{table*}[th!]
    \centering
    \begin{tabular}{|c|c|c|c|c|}
         \hline Experiment & Material & Dimensions & $\lambda_\mathrm{max}$ & $m_{\rm DM}$ cutoff  \\
         &  & [cm] & [cm] & [keV]  \\\hline\hline
         \multicolumn{5}{|c|}{Past and current experiments}\\\hline\hline
        EDELWEISS III~\cite{Armengaud_2018} & Ge & H: 4, $\varnothing$: 7 & 2.2 & 500 \\
        SuperCDMS Soudan~\cite{Aralis_2020} & Ge & H: 2.5, $\varnothing$: 7.6 & 2.2  & 500 \\
        GERDA (HPGe) \cite{Agostini_2020} & Ge & H: 7--11, $\varnothing$: 6--8$^\dagger$ & 2.6 & 1000 \\
        GERDA (BEGe) \cite{Agostini_2020} & Ge & H: 2.5--5, $\varnothing$: 6.5--8 & 2.6 & 1000 \\
        XENON1T~\cite{Aprile_2020} & Xe & H: 97, $\varnothing$: 96 & 0.88 & 200 \\
        PandaX-4T~\cite{meng2021dark} & Xe & H: 130, $\varnothing$: 100 & 4.2 & 1000 \\\hline\hline
         \multicolumn{5}{|c|}{Upcoming experiments}\\\hline\hline
        SuperCDMS SNOLAB \cite{Bloch_2017,Agnese_2017} & Si & H: 3.3, $\varnothing$: 10 & 2.1 & 100* \\ 
        SuperCDMS SNOLAB \cite{Bloch_2017,Agnese_2017} & Ge & H: 3.3, $\varnothing$: 10 & 0.3 & 100* \\ 
        LZ~\cite{thelzcollaboration2021projected} & Xe & H: 150, $\varnothing$: 150 & 0.09 & 85 \\
        DARWIN \cite{Aalbers_2016,Redard-Jacot:2020cwn} & Xe & H: 260, $\varnothing$: 260 & 4.2  & 1000  \\\hline
    \end{tabular}
    \caption{Relevant details of current and upcoming experiments with sensitivity to dark bosons. % published dark photon limits or projections. 
    The highest DM mass that can be probed is referred to as $m_\mathrm{DM}$ cutoff. The respective maximum attenuation length, calculated at the corresponding outgoing photon energy, is given by $\lambda_\mathrm{max}$. H and $\varnothing$ refer to the height and diameter of the time-projection chamber (Xe) or crystal target (Si, Ge). * The cutoff for SuperCDMS SNOLAB is a conservative estimate based on the dynamic range demonstrated in Ref.~\cite{Ren_2021}. $^\dagger$ The GERDA HPGe detectors have a central, $\sim 1$\,cm wide bore hole reducing the effective diameter accordingly. } 
    \label{tab:exp}
\end{table*}

The stage is now set to include dark Compton scattering into what would typically be considered dark absorption searches. As shown above, the expected signal for dark Compton scattering is a delta function in electron recoil energy at $T_{\rm C}=\frac{m_{\rm DM}^2}{2(m_e+m_{\rm DM})}$ with $m_{\rm DM}$ the mass of the dark boson, and is thus expected at an energy that is lower than the expected dark boson absorption signal at $T_{\rm A}=m_{\rm DM}$. Both signals would appear as a line, or as a peak if detector resolution effects are taken into account, at energy $T_{\rm C}$ or $T_{\rm A}$ respectively. 
Using the rates for dark Compton scattering, Eqs.~\eqref{eq:Rate_DP_long} and \eqref{eq:Rate_ALP_long}, and those for absorption, Eqs.~\eqref{equ:AbsRate_DP} and~\eqref{equ:AbsRate_ALPs},  exclusion limits and reach into DM parameter space can be computed for any target material of interest. 

To demonstrate the individual contributions of each process to the reach we assume zero backgrounds during an exposure of 1\,kg-year and neglect detector effects such as signal efficiency, energy resolution and thresholds. We further assume a single interaction in an event within the target material. The resulting projected reaches at 90\%\,C.L. are shown in Fig.~\ref{fig:zero_bkg_limits} for some of the most common target materials in direct DM searches. Substantially enhanced reach into DM parameter space is possible at high DM masses compared to previous estimates when including the previously neglected dark Compton scattering as additional DM interaction channel. The crossover in reach at around 100\,keV is, just like within the Standard Model (SM), due to an increased probability for Compton scattering over absorption at higher energies. Overall the dark Compton scattering bounds are less dependent on the target material than the respective dark absorption bounds. The only target-dependent quantity entering the rates Eqs.~(\ref{eq:Rate_DP_long}) and (\ref{eq:Rate_ALP_long}) is the electron density, which is very similar for typical direct DM search materials. However exceptions with a notable higher electron density do exist, such as diamond.

When a dark boson undergoes dark Compton scattering, resulting in an emitted SM photon, the above assumption of a single interaction per event in the detector is often inaccurate. Instead, secondary interactions of the emitted SM photon may occur in the detector. A key parameter is the ratio of the diameter or thickness $d$ of the detector to the photon attenuation length $\lambda$, which impacts what happens after the initial DM-detector interaction. Table~\ref{tab:exp} collects relevant scales for various existing and proposed direct DM detection and low background neutrino experiments with sensitivity to dark boson masses $m_{\rm DM} \gtrsim 100$\,keV, where dark Compton scattering dominates.

{\bf Thin detector.} 
In a thin detector, with $d \ll \lambda$, the outgoing SM photon produced by dark Compton scattering will leave the target material without further interaction. The total energy deposited into the detector via the recoiling electron is given by $T_\mathrm{C}$ which is smaller than the mass of the incoming dark boson $m_{\rm DM}$. 
For lower dark boson masses, the probability to be absorbed is larger than the probability to dark Compton scatter, but for higher masses (typically $m_{\rm DM}\gtrsim 10-100$\,keV) the situation is reversed.  Accordingly, in a dark absorption plus Compton search two peaks are expected to appear at different recoil energies, $T_\mathrm{A}$ and $T_\mathrm{C}$, where the relative height of the peaks scales with the respective difference in cross sections at the incoming DM mass. This has three notable consequences. First, the experiment is now sensitive to DM masses for which the absorption peak would lie beyond the dynamic range of the detector at high energies; this holds true as long as the dark Compton peak still resides within the dynamic range of the detector. Second, in a search with discovery potential the observation of a single peak can only result in a discovery claim at masses for which no significant dark Compton peak is expected within the dynamic range of the detector. Third, the {\it look elsewhere} effect \cite{Lyons_2008} is expected to be notably smaller in a discovery search for masses with two expected, correlated peaks.

{\bf Thick detector.} 
If the detector is thick, with $d \gg \lambda$, the outgoing SM photon produced by dark Compton scattering will be fully reabsorbed by the detector. The total energy deposited into the detector coming from the primary electron recoil and the reabsorbed outgoing photon is equal to $T_{\rm A} = m_{\rm DM}$, namely the same as if  the dark boson itself was completely absorbed.  Whether or not the individual energy depositions can be resolved in time or in position highly depends on the target material and detector design. For a separation in time, a timing resolution of $\mathcal{O}$(10\,psec\;--\,1\,nsec) is required. 
For a separation in space, a sub-cm spatial resolution is required. 
If the multiple interaction sites cannot be resolved, effectively one has a ``sum event'' with a total electron recoil energy of $m_\mathrm{DM}$. 
The spectral shape of the dark boson signal in an absorption plus dark Compton search would thus be exactly the same as in a pure absorption search, but with the crucial difference that the total expected signal rate would be the sum of both the absorption and the dark Compton scattering rate, thus strongly enhancing the detector reach in mass regions in which the dark Compton scattering process dominates.

\begin{figure}
    \centering
    \includegraphics[width=\columnwidth]{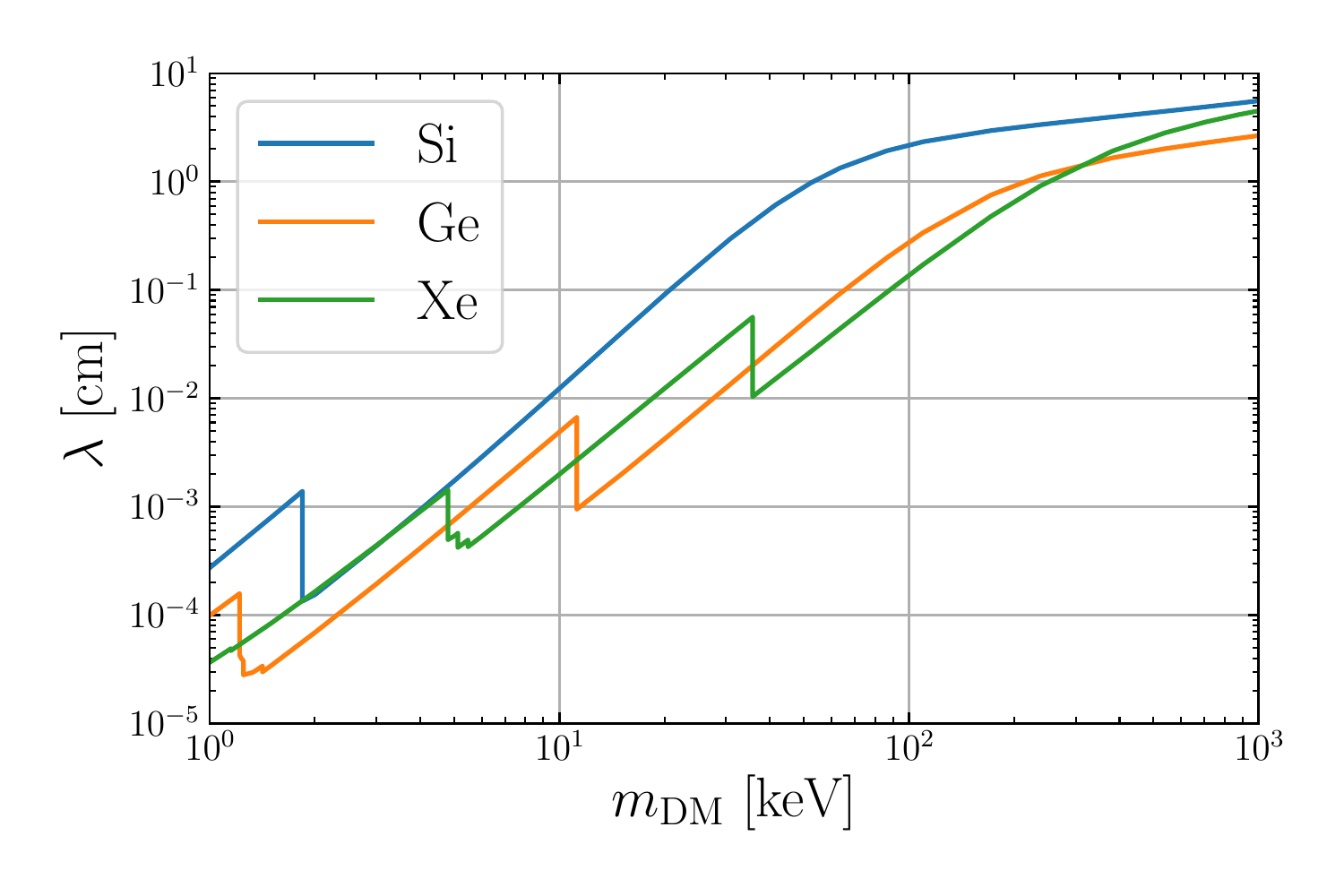}
    \caption{Attenuation length $\lambda$ for the photon emitted by a dark Compton-like process as a function of DM mass in various materials \cite{hubbell1995tables}. For $m_{\rm DM}\gtrsim 100$ keV, where the Compton-like process begins to dominate, $\lambda$ is typically of order ${\cal O}({\rm few\ cm})$. }
    \label{fig:lambda}
\end{figure}

{\bf Intermediate Detector.} If a dark boson enters an intermediate detector, with $d \approx \lambda$, here too it is either directly absorbed via dark absorption or undergoes dark Compton scattering. For the latter, there is a nonzero probability that the outgoing SM photon leaves the detector before all of its energy is lost via secondary interactions. Thus the total energy deposited lies somewhere between $T_{\rm C}$ and $T_{\rm A}$. An accurate signal model then requires the exact knowledge of the detector geometry and is best determined via simulations. Without simulations, a conservative assumption can be made based on the probability
\begin{equation}
    \label{eq:beer}
    P(x)=e^{-x/\lambda}
\end{equation}
of observing a photon, subjected to absorption and scattering processes, at depth $x$ into the target material. Here $\lambda$ is the material and energy dependent attenuation length, shown for different materials in Fig.~\ref{fig:lambda}. Using Eq.~\eqref{eq:beer} and half of the smallest detector thickness $d$ in any direction, we can calculate the minimal probability for which the outgoing SM photon would be fully absorbed within the detector. In other words, we can calculate the reduction in the rate of full absorption of the energy from the incoming dark boson, compared to e.g. the case for a thick detector. We only use one-half of the detector thickness because the incoming dark boson can interact anywhere in the detector and on average it will, to a good approximation, interact in the center of the detector. The outgoing SM photon thus only sees, on average, half of the detector. Using only the region around $T_A$ as the region of interest, a conservative limit can thus be placed in the same fashion as for thick detectors but with a suppressed rate.

Finally we note that the above zero-background toy experiments and the corresponding projected reach presented in Fig.~\ref{fig:zero_bkg_limits} are independent of the detector thickness. As long as the energies are within the dynamic range of the detector, the exact total recoil energy of the signal events matters only in the presence of backgrounds.

%%%%%%%%%%%%%%%%%%%%%%%%%%%%%%%%%%%%%%%%%%%%%%%%%%%%%%%%%%%%%%%%%
\begin{figure*}[th!]
    \centering
    \includegraphics[width=0.49\textwidth]{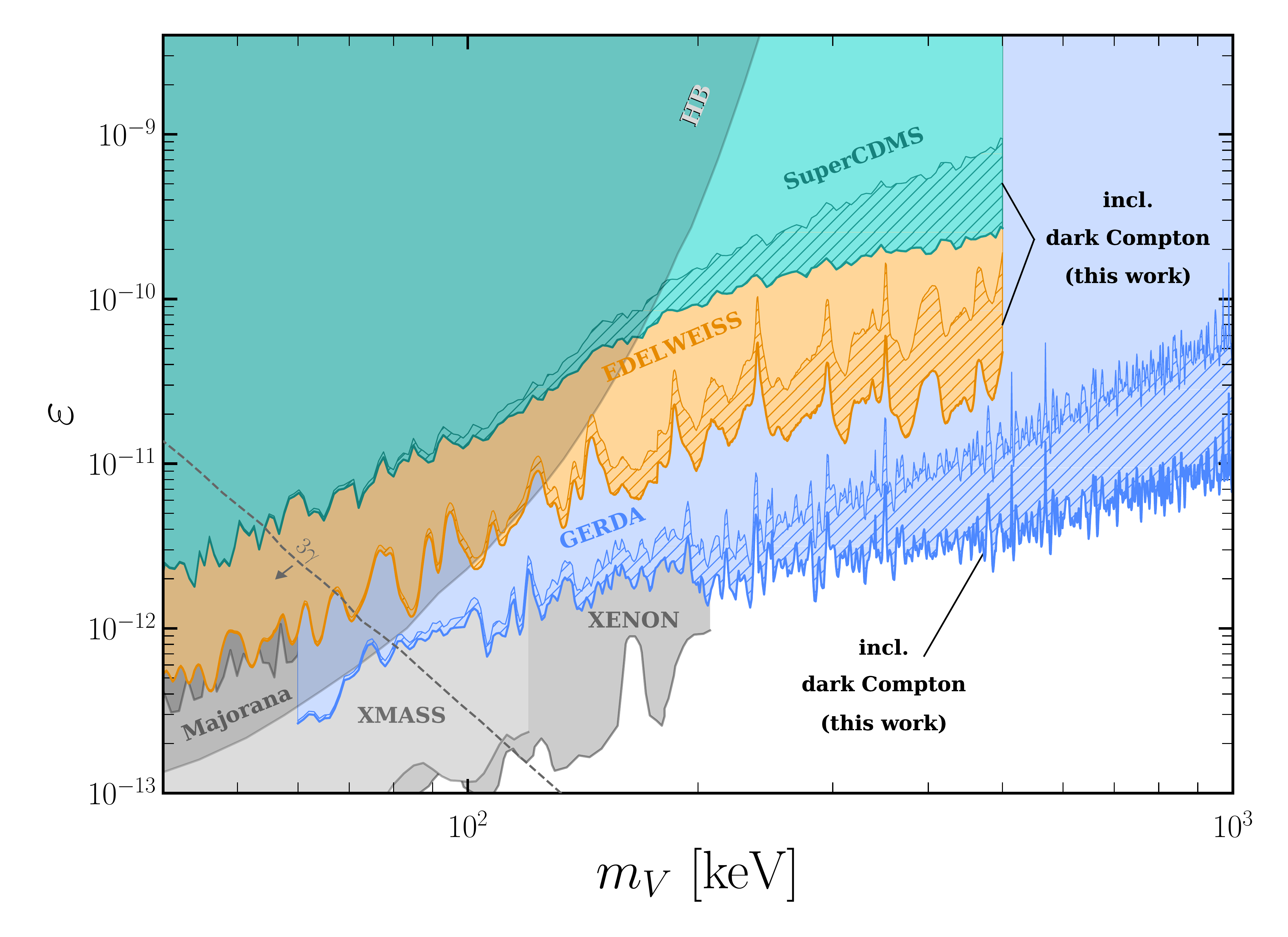}
    \includegraphics[width=0.49\textwidth]{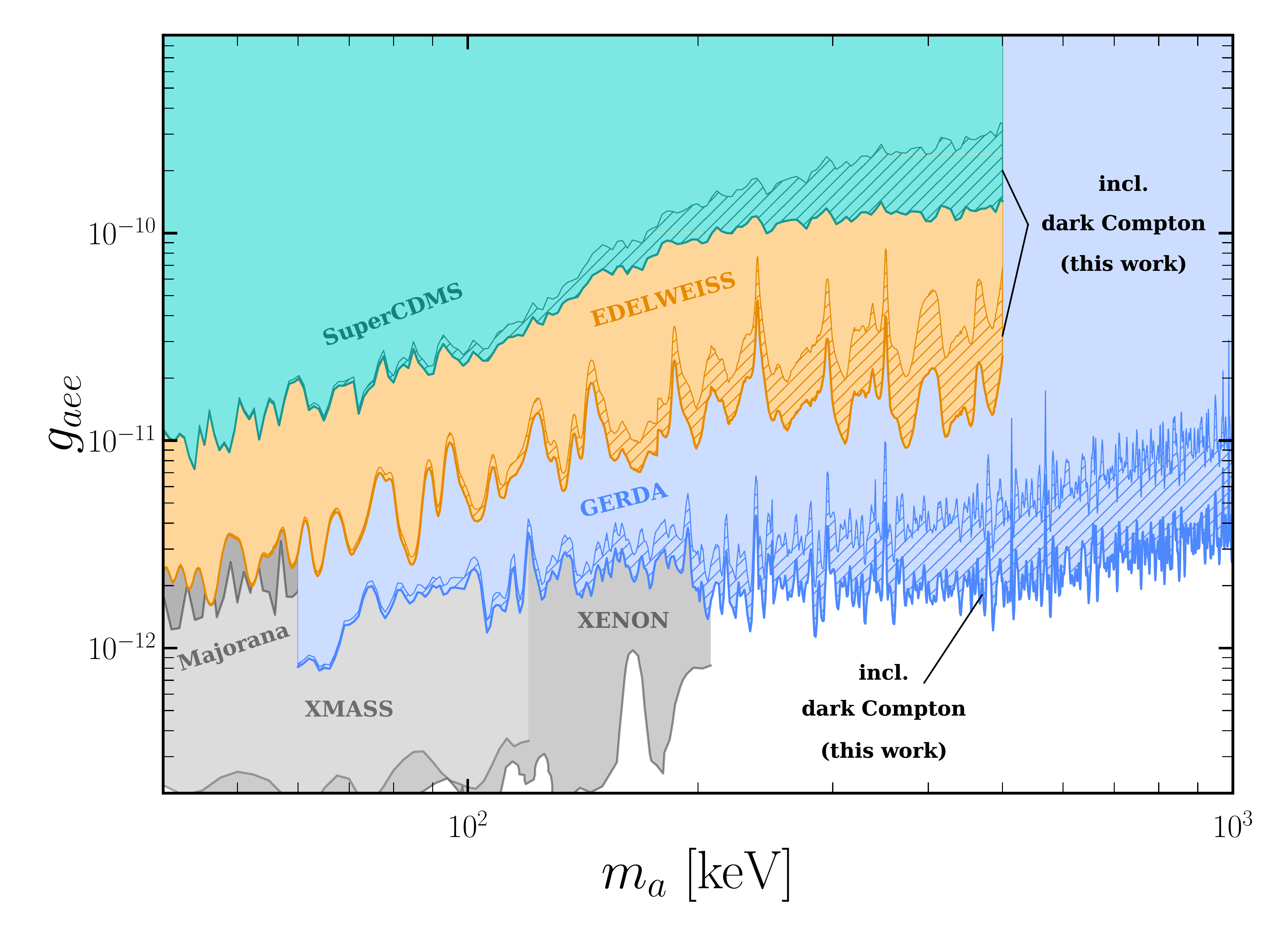}
    \caption{Constraints at 90\% C.L. on the kinetic mixing of dark photons ({\it left}) and the effective coupling of axion-like particles to electrons ({\it right}). The limits recalculated in this work for SuperCDMS Soudan \cite{Aralis:2019nfa}, EDELWEISS III \cite{Armengaud_2018} and GERDA \cite{Agostini_2020} to include dark Compton scattering (hatched) are compared to the original limits only considering dark absorption (unhatched). We also show results from XENON1T \cite{Aprile_2020}, XMASS \cite{Abe_2018}, and MAJORANA DEMONSTRATOR \cite{Abgrall_2017}, as well as constraints on the kinetic mixing parameter based on horizontal branch (HB) stars \cite{Redondo_2013} and on the diffuse gamma-ray background that limits the decay rate of dark photons into three photons ($3\gamma$) \cite{Redondo_2009}. All direct DM search results assume a local DM mass density of 0.3\,GeV/cm$^3$.}
    \label{fig:Limit_recasts}
\end{figure*}

\section{Impact on Existing Limits}

In typical direct DM search experiments and in various low background neutrino experiments, both dark absorption and dark Compton events are expected to be observed as electron recoil events. At the recoil energies of interest in this {\it Letter}, $\mathcal{O}$(keV--MeV), electron recoil searches are usually dominated by radioactive and cosmogenic backgrounds against which DM events cannot be discriminated on an event-by-event basis \cite{Aralis:2019nfa,Armengaud_2018,Agostini_2020,Aprile_2020,Abe_2018,Abgrall_2017}. Instead, a time integrated electron recoil energy spectrum after a certain exposure is analyzed on a statistical basis. Respective exclusion limits resulting from direct DM absorption searches are shown in Fig.~\ref{fig:Limit_recasts}, representing the current landscape for dark photons and ALPs. These searches are limited to dark boson masses up to twice the electron mass at $\sim 1$\,MeV in order to maintain stability of the DM from decays into electron-positron pairs.

Three experiments, SuperCDMS~\cite{Aralis:2019nfa}, EDELWEISS~\cite{Armengaud_2018}, and GERDA~\cite{Agostini_2020}, have demonstrated sensitivity to dark boson masses notably above $100$\,keV and thus into the mass region at which dark Compton processes are expected to dominate. We therefore chose these three experiments to recalculate the previously published results to include dark Compton scattering in the expected signal event rate. All three experiments are germanium based and qualify as {\it intermediate} detectors as discussed above, given the target dimensions summarized in Tab.~\ref{tab:exp}. To calculate the attenuated dark Compton rate of fully absorbed outgoing photons, we use half of the smallest dimension of each detector. For SuperCDMS and EDELWEISS, an effective target thickness of 1.25\,cm and 2\,cm, respectively, is assumed in any direction. For GERDA, an effective target thickness of 1.25\,cm is assumed for the 30 BEGe detectors and of 2.5\,cm for the 7 HPGe detectors, taking the 1\,cm bore hole in the HPGe detectors into account. The strongest attenuation is expected for GERDA BEGe detectors with $d=1.25$\,cm and $\lambda_\mathrm{max}=2.6$\,cm at $m_\mathrm{DM}=1$\,MeV. Using Eq.~\eqref{eq:beer}, we find the attenuated full-absorption Compton rate in this example to be $\sim 38$\% of the original dark Compton rate. All other detectors (GERDA HPGe, SuperCDMS, and EDELWEISS) and masses considered have higher rate fractions up to 100\% in GERDA HPGe detectors at $m_\mathrm{DM}=100$\,keV.

For the new limits we present,  we consider only events in which the incoming energy has been fully absorbed when determining the expected DM event rate. This approach 
artificially sets the DM signal efficiency for all other events to zero, while the background efficiency remains unchanged, yielding a conservative bound.  The resulting updated limits for SuperCDMS, EDELWEISS, and GERDA are presented in Fig.~\ref{fig:Limit_recasts}. Above $\sim 100$\,keV masses, the reach of all three experiments increases with mass compared to the previously published absorption-only searches, due to the increasing probability for dark Compton scattering over absorption. When including dark Compton scattering, GERDA is able to cover previously uncharted parameter space, improving its current reach at 1\,MeV by a factor of $\sim 3$ for $g_{aee}$ and $\sim 6$ for $\varepsilon$. Further improvement can be achieved with an accurate treatment of the DM signal efficiency  
for nonfully absorbed events and when the full detector geometry is taken into account, rather than our treatment of the BEGe and HPGe detectors as spheres with radii 1.25\,cm and 2.5\,cm  respectively.

%%%%%%%%%%%%%
\section{Outlook}

In this work we have shown that direct detection experiments searching for DM in the laboratory are substantially more powerful at probing high-mass dark relic bosons than previously thought. Dark Compton scattering that can occur in the detector becomes prominent at high masses above $\gtrsim 100\; {\rm keV}$ and their inclusion in the analysis for both limit-setting and discovery carries significant impact. This is demonstrated by our estimate for the improved projected reach in a zero-background setting, as well as in the conservative recalculation of existing limits from SuperCDMS, EDELWEISS and GERDA results. Our work advocates for the simulation of the detector geometry to make use of the full target volume, to build a complete dark Compton scattering signal model (including corrections from atomic energy levels and any potential many-body effects) over all accessible electron recoil energies and to maximize the respective signal efficiency. In doing so, even stronger constraints and reach than presented in this work are expected.
%This is to list the suppl. material reference also in the main paper:
\nocite{1994okml.book..113K}

%%%%%%%%%%%%%
\bigskip
\begin{acknowledgments}
{\bf Acknowledgements.} We thank (in alphabetical order) Eve Bodenia, Paul Brink, Torben Ferber, Camilo Garcia-Cely, Sunil Golwala, Sam McDermott,  Scott Oser, Richard Partridge and Hadar Steinberg for useful discussions. We are grateful to Rouven Essig, \mbox{Torben} Ferber, and Noah Kurinsky for comments on a draft version of this manuscript. 
% We thank Camilo Garcia-Cely for useful conversations related to the content of this paper. 
The work of Y.H. is supported by the Israel Science Foundation (Grant No. 1112/17), by the Binational Science Foundation (Grant No. 2016155), by the I-CORE Program of the Planning Budgeting Committee (Grant No. 1937/12),  and  by the Azrieli Foundation. The work of B.v.K. is funded by the Deutsche Forschungsgemeinschaft (DFG) %, German Research Foundation) 
under the Emmy Noether Grant No. 420484612, and under Germany's Excellence Strategy - EXC 2121,  390833306. The work of E.K. is supported by the Israel Science Foundation (Grant No.1111/17), by the Binational Science
Foundation (Grant No. 2016153), and by the I-CORE
Program of the Planning Budgeting Committee (Grant
No. 1937/12).  T.C.Y. is supported by the U.S. Department of Energy under Contract Number DE-AC02-76SF00515.
\end{acknowledgments}

% !!! The appendix is only activated for the arXiv submission - the journal version uses suppl. material !!!
\section{Appendix: Derivation of Dark Compton Rates}
\label{apdx:rates}

{\bf Dark Photons.} The differential cross-section for dark Compton scattering is given by
\begin{align}
    d\sigma &= \frac{1}{4m_e\omega\; v_{\rm rel}} d\Pi\; \langle |\mathcal{M}|^2 \rangle 
\end{align}
where $\omega$ is the energy of the incoming dark photon $V$, or more generally speaking of the incoming dark boson. $m_e$ is the mass of the electron, $v_{\rm rel}$ is the velocity of the Earth relative to the galactic rest frame and

\begin{align}
\int d\Pi 
    &=  \int \frac{d\cos\theta}{8\pi} \frac{\omega'}{\omega + m_e - \sqrt{\omega^2-m_V^2}\cos\theta}\,.
\end{align}

\noindent{}In this equation, $m_V$ is the mass of the dark photon and $\theta$ and $\omega'$ are the scattering angle and the energy of the outgoing Standard Model (SM) photon, respectively. For dark photons the matrix element is given by

\begin{align}
    \langle|\mathcal{M}|^2 \rangle &= \frac{e^4\varepsilon^2}{6} \left(\frac{2m_e^2+m_V^2}{\omega'^2}+\frac{4m_e\omega'}{2m_e\omega+m_V^2}\right. \nonumber\\ \nonumber
    &+ \left.\frac{4(2m_e^2+m_V^2)(m_e^2+2m_e\omega+m_V^2)}{(2m_e\omega+m_V^2)^2}\right.\\ 
   &+ \left.\frac{m_V^4+4m_e^2(\omega^2-m_V^2)-8m_e^3(\omega+m_e)}{m_e\omega'(2m_e\omega+m_V^2)} \right)\,.
\end{align}

\noindent{}$\varepsilon$ is the kinetic mixing strength of the dark photon with ${\cal L}\supset \frac{\varepsilon}{2}F_{\mu\nu}  F'^{\mu\nu}$ and with $F_{\mu\nu}\ (F'_{\mu\nu}$) being the (dark) photon field strength. We do not include in-medium effects as they are insignificant in the high DM mass regime~\cite{Hochberg:2016sqx} where dark Compton scattering plays a meaningful role.

Converting the cross-section to an event rate (per unit time per unit mass) via
\begin{align}
    R = \frac{1}{\rho_{\rm T}}\frac{\rho_{\rm DM}}{m_{V}} n_e \sigma v_{\rm rel}\,,
\end{align}

\noindent{}with the DM mass density $\rho_{\rm DM}$, the number density of electrons in the target $n_e$ and the mass density of the target $\rho_{\rm T}$,
we have

\begin{align}
        \frac{dR}{d\cos\theta} &= \frac{n_e}{\rho_{\rm T}}\frac{\rho_{\rm DM}}{m_{V}} \frac{e^4\varepsilon^2\omega'}{48\pi m_e\omega (\omega + m_e - \sqrt{\omega^2-m_V^2}\cos\theta) }\times \nonumber \\ \nonumber
    &\left(\frac{2m_e^2+m_V^2}{\omega'^2}+\frac{4m_e\omega'}{2m_e\omega+m_V^2}\right. \\ \nonumber
    &+ \left.\frac{4(2m_e^2+m_V^2)(m_e^2+2m_e\omega+m_V^2)}{(2m_e\omega+m_V^2)^2}\right.\\ 
   &+ \left.\frac{m_V^4+4m_e^2(\omega^2-m_V^2)-8m_e^3(\omega+m_e)}{m_e\omega'(2m_e\omega+m_V^2)} \right)\,. \label{eq:darkphoton_rate}
\end{align}

Note that this expression does not exactly reduce to the Klein-Nishina formula \cite{1994okml.book..113K} for SM photons in the $m_V\to0$ limit since the dark photon, being a massive vector boson, has one more degree of freedom than the SM photon.

The total dark Compton rate for a slow-moving DM flux is found through integration and setting $\omega\approx m_V$, yielding
\begin{align}
    \label{eq:Rate_DP_long_appendix}
    R &= \frac{n_e}{\rho_{\rm T}}\frac{\rho_{\rm DM}}{m_{V}} \frac{e^4\varepsilon^2}{24\pi m_e^2}\frac{ (m_V+2m_e) (m_V^2+2m_em_V+2m_e^2) }{(m_V + m_e)^3}  \\
     &\approx \frac{n_e}{\rho_{\rm T}}\frac{\rho_{\rm DM}}{m_{V}} \frac{e^4\varepsilon^2 }{24\pi m_e^2} \times \left\{
    \begin{array}{ccc}
      1  & & {\rm for}\ m_V \gg m_e \\
        4 & & {\rm for}\ m_V \ll m_e\\
    \end{array}\right.
\end{align}

{\bf Axion-like Particles.} Next consider an ALP, $a$, that couples to electrons via the interaction 
\begin{align}
\mathcal{L}_{aee}=  {g_{aee}}i a \bar{\psi}_e \gamma^5 \psi_{e} 
\end{align}

\noindent{}and with a coupling parameter $g_{aee}$ that undergoes dark Compton scattering. Being a pseudo-scalar the ALP has only one degree of freedom. The matrix element is given by

\begin{align}
    \langle|\mathcal{M}|^2 \rangle &= \frac{e^2g_{aee}^2}{4} \left(\frac{m_a^2}{\omega'^2}+\frac{m_a^4-4m_a^2m_e^2+4m_e^2\omega^2}{m_e\omega'(m_a^2+2m_e\omega)}\right. \nonumber\\ 
    &+ \left.\frac{4m_a^2m_e^2}{(m_a^2+2m_e\omega)^2}+\frac{4m_e\omega'-8m_e\omega}{m_a^2+2m_e\omega}\right)\,,
\end{align}

\noindent{}and so the differential rate is 

\begin{align}
\frac{dR}{d\cos\theta} &= \frac{n_e}{\rho_{\rm T}}\frac{\rho_{\rm DM}}{m_{a}} \frac{e^2 g^2_{aee}\omega'}{32\pi m_e\omega(\omega+m_e-\sqrt{\omega^2-m_a^2}\cos\theta)}\times \nonumber \\
    & \left(
    \frac{m_a^4-4m_a^2m_e^2+4m_e^2(\omega-\omega')^2}{m_e\omega'(m_a^2+2m_e\omega)} \right. \nonumber \\  &\left. +\frac{m_a^2}{\omega'^2}+\frac{4m_a^2m_e^2}{(m_a^2+2m_e\omega)^2}\right)\,
\end{align}

\noindent{}with $m_a$ being the ALP mass. For a slow-moving DM flux, we have 

\begin{align}
\label{eq:Rate_ALP_long_appendix}
R &= \frac{n_e \rho_{\rm DM}}{\rho_{\rm T}}\frac{e^2 g^2_{aee} m_a (m_a+2m_e)^2}{ 16\pi m_e^2(m_a+m_e)^4} \\
&\approx \frac{n_e}{\rho_{\rm T}}\frac{\rho_{\rm DM}}{m_a}\frac{e^2 g^2_{aee} }{ 16\pi m_e^2} \times \left\{ \begin{array}{ccc}
      1  & & {\rm for}\ m_a \gg m_e \\
        4m_a^2/m_e^2 & & {\rm for}\ m_a \ll m_e\\
    \end{array}\right.
\end{align}
Interestingly, this rate is maximized at $m_a \approx 0.56 m_e$.

\clearpage

\bibliography{main}

\clearpage

\end{document}